\documentclass[prd,aps,a4paper,nofootinbib,eqsecnum,twocolumn]{revtex4}  

\newif\ifusesec
\usesectrue  
   
\usepackage{graphicx} 
\usepackage{mathrsfs}
\usepackage{amsmath,amsfonts,amssymb}
\usepackage{multirow}

\newcommand{\beq}{\begin{equation}}
\newcommand{\eeq}{\end{equation}}
\newcommand{\bea}{\begin{eqnarray}}
\newcommand{\eea}{\end{eqnarray}}

\begin{document}

\title{Investigating new forms of gravity-matter couplings in the gravitational field equations}

\author{Donato \surname{Bini}$^{1,2}$, Giampiero Esposito$^{3,2}$}

\affiliation{$^1$Istituto per le Applicazioni del Calcolo ``M. Picone'', CNR, I-00185 Rome, Italy\\
Orcid: 0000-0002-5237-769X}
\affiliation{$^2$Istituto Nazionale di Fisica Nucleare, \\
Complesso Universitario di Monte S. Angelo,
Via Cintia Edificio 6, 80126 Napoli, Italy}
\affiliation{$^3$ Dipartimento di Fisica ``Ettore Pancini'', \\
Complesso Universitario di Monte S. Angelo,
Via Cintia Edificio 6, 80126 Napoli, Italy\\
Orcid: 0000-0001-5930-8366} 

\begin{abstract}
This paper proposes a toy model where, in the Einstein equations, 
the right-hand side is modified by the addition of a term
proportional to the symmetrized partial contraction of the Ricci
tensor with the energy-momentum tensor, while the left-hand
side remains equal to the Einstein tensor. Bearing in mind the
existence of a natural length scale given by the Planck length,
dimensional analysis shows that such a term yields a correction
linear in $\hbar$ to the classical term, that is instead just proportional
to the energy-momentum tensor. One then obtains an effective
energy-momentum tensor that consists of three contributions: 
pure energy part, mechanical stress and thermal part. 
The pure energy part has the appropriate property for dealing
with the dark sector of modern relativistic cosmology.
Such a theory coincides with
general relativity in vacuum, and the resulting field equations
are here solved for a Dunn and Tupper metric, for departures from 
an interior Schwarzschild solution as well as for a 
Friedmann-Lemaitre-Robertson-Walker universe.
\end{abstract}

\date{\today}

\maketitle

\section{Introduction}

At the time when Einstein assumed that gravity couples to the
energy-momentum tensor of matter, it was not yet known that 
matter fields are quantum fields in the first place, and no
attempt had been made to understand the physical implications
of the Planck length
\begin{equation}
\ell_P \equiv \sqrt{G \hbar \over c^{3}}.
\label{(1.1)}
\end{equation}
Einstein obtained his field equations in the well-known form
\cite{E1915,E1916}
\begin{equation}
E_{\mu\nu}\equiv R_{\mu \nu}-{1 \over 2}g_{\mu \nu}R+\Lambda g_{\mu\nu}
= \kappa  T_{\mu \nu},\qquad \kappa={8 \pi G \over c^{4}},
\label{(1.2)}
\end{equation}
whose contracted covariant differentiation
leads therefore to the local relation
\begin{equation}
\nabla^{\mu}T_{\mu \nu}=0,
\label{(1.3)}
\end{equation}
which however does not yield any integral conservation law unless
the spacetime manifold $(M,g)$ admits Killing vector fields, which
is not necessarily the case in a generic spacetime.

When the renaissance of general relativity and cosmology 
\cite{Ellis} began in
the sixties, several approaches were developed along the years
in order to modify the classical field equations \eqref{(1.2)}:
\vskip 0.3cm
\noindent
(i) Quantum field theory in curved space-time
\cite{DeWitt1975,Birrell,Fulling,Buchbinder,Kay,PToms}, where the classical
energy-momentum tensor is replaced by the expectation value
$\langle T_{\mu \nu} \rangle$ of its regularized and
renormalized form in a quantum state (the choice of
quantum state being taken not to affect the result). With the help of
point-split regularization or heat-kernel methods, one can
therefore obtain a number of correction terms quadratic in the
curvature \cite{Birrell,Vilko,Parker}. This is certainly relevant as
one approaches the quantum era, which affects the very early universe.
\vskip 0.3cm
\noindent
(ii) Full quantum gravity via functional integrals is studied
\cite{DW1,DW2,DW3}, writing down the functional equations obeyed by
the effective action, possibly allowing for supersymmetry and 
supergravity \cite{Ferrara,SUGRA,Freedman}.
\vskip 0.3cm
\noindent
(iii) One resorts to string theory where, at perturbative level,
spacetime is described by a set of coupling constants in a
two-dimensional quantum field theory, whereas at non-perturbative
level spacetime must be reconstructed from a holographic dual theory
\cite{Horowitz}.
\vskip 0.3cm
\noindent
(iv) One studies instead the most general family of classical, 
relativistic Lagrangians for the gravitational 
field \cite{GRG,IJGMMP}.

The latter has given rise to the so-called $f(R)$ theories 
\cite{fR1,fR2,fR3,fR4} and their
many variants, where the Lagrangian is no longer linear in the 
trace of the Ricci tensor. This is certainly relevant for the
analysis of classical phenomena such as the expansion of the
universe. However, it is then difficult
to develop a rigorous theory of the Cauchy problem of the
same standard of rigor now available in general relativity
\cite{CA1,CA2}. Moreover, it is unclear 
how to achieve a smooth transition towards general relativity
in the solar system, where Einstein's theory has been successfully
tested \cite{W1}, showing no compelling need for alternative 
classical theories \cite{W2}.
In other words, as far as the large-scale structure of the universe
is concerned, the discovery of the acceleration of the universe
\cite{ACC} cannot be understood by using general relativity, and
hence one resorts to alternative classical Lagrangians. But the 
smooth transition to general relativity, at least on solar-system
scale, deserves further work, as far as we can see. 

In light of all the above well known properties or open problems,
we have been led to consider a modification of Eqs. \eqref{(1.2)} 
that fulfills the following requirements:
\begin{enumerate}

\item The Einstein-Hilbert Lagrangian is not modified.

\item The modified theory coincides with general relativity when
the energy-momentum tensor source of the gravitational field  vanishes.
The modified field equations correct only the right-hand side
of Eqs. \eqref{(1.2)}, by means of an additional term that is a 
symmetrized partial contraction of the Ricci tensor with the
energy-momentum tensor of matter.
\end{enumerate}

Because of these goals, we assume a tensor equation reading as
(we use summation over repeated indices, and the symbol
$|\alpha|$ to denote that the index $\alpha$ among the others is not 
affected by symmetrization over its adjacent index) 
\begin{equation}
E_{\mu\nu}
=\kappa 
T_{\mu \nu} + B R_{(\mu}^{\; \; \; \alpha} 
\; T_{|\alpha|\nu)}.
\label{(1.4)}
\end{equation}
Note that the left-hand side is classical and results from variation
of the Einstein-Hilbert action, while the right-hand side is tensorial
but phenomenological, since it is affected by possible quantum
laws of coupling gravity to matter fields.
The coefficient $B$ should be therefore dimensionful, and in such a
way that the dimension of $B$ times the dimension of Ricci equals the
dimension of ${G \over c^{4}}\propto \kappa$. Thus, bearing in mind that the
length scale is set by the Planck length \cite{Parker},
on denoting by $b$ an arbitrary real number, we write
\begin{equation}
B = b \kappa (\ell_P)^{2},  
\label{(1.5)}
\end{equation}
as well as
\begin{equation} 
A=\frac{B}{\kappa}=b (\ell_P)^{2},
\label{(1.5bis)}
\end{equation}
where we have introduced the related  quantities $b$ (dimensionless) and 
$A$ (with the dimension of a length squared) for convenience and a later 
use. Both these quantities can be either positive or negative.
Hence Eq. \eqref{(1.4)} can be re-expressed in the form
\begin{equation}
E_{\mu\nu} =\kappa 
\left[T_{\mu \nu}+A
R_{(\mu}^{\; \; \; \alpha} \;
T_{|\alpha|\nu)}\right].
\label{(1.6)}
\end{equation}

Let us note in passing that the cosmological constant term $\Lambda$ 
can be equivalently shifted from one side to the other of the Einstein's 
equations and included for instance in the energy-momentum tensor 
source of the spacetime curvature as a $T^\Lambda_{\mu\nu}=
-\frac{\Lambda}{\kappa} g_{\mu\nu}$. We will not follow this 
shift here, using the notation $E_{\alpha\beta}$ for a full Einstein 
tensor including $\Lambda$, i.e., 
$E_{\alpha\beta}=G_{\alpha\beta}+\Lambda g_{\alpha\beta}$.
Furthermore, at this stage, nothing can be said about the 
real-valued parameter $b$, but Eq. \eqref{(1.6)} 
tells us that the classical Einstein equations \eqref{(1.2)} can be viewed 
as the zeroth-order in $\hbar$ of a richer scheme.\footnote{We note
incidentally that, in a rather different context, the full contraction
of the Ricci tensor with the energy-momentum tensor (whereas we
consider their partial contraction in Eq. \eqref{(1.6)}) is met in quantum 
Yang-Mills theory. Indeed, as pointed out in Ref. \cite{DW3}, the
presence of $m_{P}^{-2} \; R_{\rho \nu}T^{\rho \nu}$ in the $a_{2}$
heat-kernel coefficient means that, although the Yang-Mills coupling
constant gets renormalized, the finite part of the effective action
now depends on the auxiliary mass in a way that cannot be absorbed 
into a running coupling constant. Each choice of auxiliary mass 
corresponds to a different theory. Thus, the coupling to the
gravitational field destroys the perturbative renormalizability of
the Yang-Mills field, even in the purely Yang-Mills sector.}
By virtue of the arbitrariness of $b$, it appears desirable to work
with finite values of $b$, without imposing the limit as
$b$ approaches $0$.

Obviously, the proposed modification/extension of the field equations is just one of the many 
conceivable modifications. In fact, along the same lines, one could have equally 
considered coupling terms of the type (cf. Appendix A)
$$
RT \, g_{\mu\nu} ,\qquad R T_{\mu\nu},
\qquad E_{(\mu}^{\; \; \alpha} \; T_{|\alpha|\nu )},\ldots
$$
This circumstance (i.e., the possibility to consider other choices) 
is not crucial in the present study. In fact, we are only  
interested here in analyzing the consequences of one of the choices in such a family. 
It is also worth noting that we are not changing the gravitational Lagrangian,
at the price of introducing the coupling
$$
b {8\pi G \over c^{4}} (\ell_P)^{2}
R_{(\mu}^{\; \; \alpha} \; T_{|\alpha|\nu )}.
$$
By virtue of Eq. \eqref{(1.6)} and of the Bianchi identity, the local relation
\eqref{(1.3)} is now replaced by
\begin{equation}
\nabla^{\mu}\tau_{\mu \nu}=0,
\label{(1.7)}
\end{equation}
having defined the {\it effective energy-momentum tensor}
\begin{equation}
\tau_{\mu \nu} \equiv 
T_{\mu \nu}+A R_{(\mu}^{\; \; \; \alpha}
\; T_{|\alpha|\nu)},
\label{(1.8)}
\end{equation}
which by rescaling the Ricci tensor by a natural length scale associated with it, say $L$,
\begin{equation}
R_{\mu\nu}=L^{-2} \tilde R_{\mu\nu}
\end{equation}
($\tilde R_{\mu\nu}$ dimensionless) becomes
\begin{equation}
\tau_{\mu \nu} 
=T_{\mu \nu}+\epsilon  \tilde R_{(\mu}^{\; \; \; \alpha}
\; T_{|\alpha|\nu)}, \qquad \epsilon 
= b \left(\frac{\ell_P}{L}\right)^2= \frac{A}{L^2} .
\label{(1.8bis)}
\end{equation}
In all the explicit examples studied in the rest of the paper we will 
explore the case $\epsilon \ll 1$, but evidently $\epsilon$ does not 
need to be considered small at all. 

Section II studies in detail our effective energy-momentum tensor.
Sections III, IV and V are devoted to modifications of perfect
fluid spacetimes, spherically symmetric static spacetimes sourced by a 
perfect fluid and FLRW spacetimes sourced by a perfect fluid and with
nonvanishing cosmological constant, respectively. Concluding remarks are
made in Sec. VI, and relevant technical details are provided in Appendix A.

\section{Structure of the effective energy-momentum tensor}

Let us consider matter sources of the Einstein's field equations, i.e., 
a perfect fluid described by the energy-momentum tensor
\begin{equation}
T_{\mu\nu}=(\rho +p) u_\mu u_\nu+p g_{\mu\nu}
\end{equation}
with $u$ the four-velocity vector corresponding to the rest frame of the fluid, 
$\rho$ the (proper) energy density and $p$ the proper isotropic pressure.
Let us introduce the following convenient  notation for contractions of a tensor, 
say $S$, with a vector, say $X$,
\begin{equation}
S_{X}{}_{\alpha}=X^{\beta} S_{\beta \alpha} ,
\end{equation}
as well as the standard\lq\lq 1+3" decomposition of the Ricci tensor, 
$R_{\mu\nu}$, parallel and orthogonal to $u$, 
\begin{eqnarray}
R_{\mu\nu}&=&\delta_\mu^\alpha \delta_\nu^\beta R_{\alpha\beta}
\nonumber\\
&=& [\Pi(u)_\mu{}^\alpha-u_\mu u^\alpha][\Pi(u)_\nu{}^\beta-u_\nu u^\beta]R_{\alpha\beta}
\nonumber\\
&=& [\Pi(u)R]_{\mu\nu}- \Pi(u)_\nu{}^\beta u_\mu u^\alpha R_{\alpha\beta} 
\nonumber\\
&&-\Pi(u)_\mu{}^\alpha u_\nu u^\beta R_{\alpha\beta}+
u_\mu u^\alpha u_\nu u^\beta R_{\alpha\beta}
\nonumber\\
&=& R^{\perp \perp}_{\mu\nu}+2R^{\perp  \Vert}_{(\mu} u_{\nu )} 
+R^{\Vert  \Vert} u_\mu u_\nu ,
\end{eqnarray}
where
\begin{equation}
\Pi(u)_{\mu\nu} \equiv g_{\mu\nu}+u_\mu u_\nu ,
\end{equation}
projects orthogonally onto $u$ and we have defined
\begin{eqnarray}
R^{\perp \perp}_{\mu\nu}& \equiv & [\Pi(u)R]_{\mu\nu}
= \Pi(u)_\mu{}^\alpha \Pi(u)_\nu{}^\beta R_{\alpha\beta}, 
\nonumber\\
R^{\perp  \Vert}_{\mu}& \equiv & -\Pi(u)_\mu{}^\alpha   
R_{\alpha\beta}u^\beta = -\Pi(u)_\mu{}^\alpha R_{\alpha u},  
\nonumber\\
R^{\Vert \Vert}& \equiv & u^\alpha u^\beta R_{\alpha\beta}=R_{uu} .
\end{eqnarray}
Upon inserting these splitted components into Eq. \eqref{(1.8)} one finds
\begin{eqnarray}
\tau_{\mu \nu}
&=& [ \rho  +A \rho R^{\Vert  \Vert} ]u_\mu u_\nu+p [\Pi(u)_{\mu\nu}
+A R^{\perp \perp}_{\mu\nu}]
\nonumber\\
&&-  A(p-\rho)R^{\perp  \Vert}_{(\mu}u_{\nu)},
\label{(1.14)}
\end{eqnarray}
which gives for $\tau_{\mu\nu}$ a pure energy part
\begin{equation}
[\tau_{\rm en}]_{\mu\nu}=\rho \left(1  -A R^{\Vert  \Vert} \right) 
u_\mu u_\nu\equiv \rho_{\rm eff} u_\mu u_\nu ,  
\label{(2.7)}
\end{equation}
a mechanical stress part
\begin{equation}
[\tau_{\rm mec}]_{\mu\nu}=p [\Pi(u)_{\mu\nu}+A R^{\perp \perp}_{\mu\nu}],
\label{(2.8)}
\end{equation} 
and a thermal part
\begin{equation}
[\tau_{\rm th}]_{\mu\nu}=-A(p-\rho)R^{\perp  \Vert}_{(\mu}u_{\nu)},
\label{(2.9)}
\end{equation}
so that
\begin{equation}
\tau_{\mu \nu}=[\tau_{\rm en}+\tau_{\rm mec}+\tau_{\rm th}]_{\mu\nu}.
\label{(2.10)}
\end{equation}
For the original perfect fluid energy-momentum tensor 
one could have written the equivalent expression
\begin{equation}
T_{\mu\nu}=\rho u_\mu u_\nu +p \Pi(u)_{\mu\nu} ,
\label{(2.11)}
\end{equation}
showing the absence of thermal stresses in the proper reference frame, 
coherent with the definition of perfect fluid.
Note that since there are no a priori sign restrictions 
on $R^{\Vert  \Vert}$, it is legitimate 
to expect either positive or negative values for 
$\rho_{\rm eff}$, a property which 
is of basic importance to model dark matter, dark energy or even 
exotic types of matter. Moreover, the constant $A$ can be replaced by 
$\epsilon$ if one rescales the Ricci tensor by a squared length scale.

In the case $p=0$ the mechanical stress term cancels out
\begin{equation}
\tau_{\mu \nu}=\rho_{\rm eff} u_\mu u_\nu+A \rho R^{\perp  \Vert}_{(\mu}u_{\nu)} ,
\label{(2.12)}
\end{equation}
while the thermal stress disappears only in a frame where 
$R^{\perp  \Vert}_{(\mu}u_{\nu)}$ vanishes.

Things are much simpler when using an adapted frame to $u$, i.e., such that $e_0=u$ and 
$e_a$, $a=1,2,3$, span the local rest space of $u$ (Note that while $e_0$ is 
supposed to be orthogonal to $e_a$, the latter spatial vectors are not necessarily orthonormal). 
Similarly, it is useful to introduce the standard \lq\lq 1+3" decomposition of the 
Riemann tensor into its electric (${\mathcal E}(u)_{\alpha\beta}$), 
magnetic (${\mathcal H}(u)_{\alpha\beta}$) 
and mixed (${\mathcal F}(u)_{\alpha\beta}$) parts, respectively, given by
\begin{eqnarray}
\label{elemagnparts}
{\mathcal E}(u)_{\alpha\beta}&=&R_{\alpha\mu\beta\nu}u^\mu u^\nu, 
\nonumber \\
{\mathcal H}(u)_{\alpha\beta}&=&-R^{\,*}_{\alpha\mu\beta\nu}u^\mu u^\nu, 
\nonumber \\
{\mathcal F}(u)_{\alpha\beta}&=& [{}^*R^*]_{\alpha\mu\beta\nu}u^\mu u^\nu,
\end{eqnarray}
where ${\mathcal E}(u)_{[\alpha\beta]}=0={\mathcal F}(u)_{[\alpha\beta]}$ 
and ${\mathcal H}(u)^{\alpha}{}_\alpha=0$.
${\mathcal E}(u)_{\alpha\beta}$, ${\mathcal H}(u)_{\alpha\beta}$ and 
${\mathcal F}(u)_{\alpha\beta}$ are called tidal fields.
The $20$ independent components of the Riemann tensor are then summarized by the $6$ 
independent components of the electric part (spatial and symmetric tensor), the $8$ independent 
components of the magnetic part (spatial and trace-free tensor) and the $6$ independent 
components of the mixed part (spatial and symmetric tensor).
Then, in terms of frame components, all the above spatial quantities can be written as
\begin{eqnarray}
\label{elemagnparts2}
{\mathcal E}(u)_{ab}&=&R_{a0b0}, 
\nonumber \\
{\mathcal H}(u)_{ab}&=&-R^{\,*}_{a0b0}=\frac12 \eta(u)^{cd}{}_b R_{a0cd}, 
\nonumber \\
{\mathcal F}(u)_{ab}&=& [{}^*R^*]_{a0b0}
=\frac14 \eta(u)_a{}^{cd}\eta(u)_b{}^{ef} R_{cdef},
\end{eqnarray}
and can be inverted to give
\begin{eqnarray}
\label{elemagnparts2bis}
R_{a0}{}^{cd}&=&{\mathcal H}(u)_{ab}\eta(u)^{bcd} ,
\nonumber \\
R^{abcd}&=& \eta(u)^{abr}\eta(u)^{cds}{\mathcal F}(u)_{rs} ,
\end{eqnarray}
where $\eta(u)_{abc}=u^\alpha \eta_{\alpha abc}$ 
is the unit volume (spatial) three-form.
By using these relations one has also the frame components of the 
Ricci tensor $R^\alpha{}_\beta=R^{\mu\alpha}{}_{\mu\beta}$,
\begin{eqnarray}\label{riccich7}
R^0{}_0&=& -{\mathcal E}(u)^c{}_c,
\nonumber \\
R^0{}_a&=& \eta(u)_{abc}{\mathcal H}(u)^{bc},
\nonumber \\
R^a{}_b&=& -{\mathcal E}(u)^a{}_b-{\mathcal F}(u)^a{}_b
+\delta^a{}_b{\mathcal F}(u)^c{}_c ,
\end{eqnarray}
so that the curvature scalar takes the form
\begin{equation}
R=R^0{}_0+R^a{}_a=-2({\mathcal E}(u)^c{}_c-{\mathcal F}(u)^c{}_c) .
\end{equation}
On converting into the previous language, one writes 
\begin{eqnarray}
R^{\perp \perp}_{\mu\nu}&\to& R_{ab}\nonumber\\
R^{\perp  \Vert}_{ \mu} &\to& R_{0a}\nonumber\\
R^{\Vert  \Vert} &\to& R_{00} .
\end{eqnarray}
Therefore
\begin{eqnarray}
\tau_{00}=[\tau_{\rm en}]_{00} &=& \rho \left(1  -A R_{00} \right),   
\nonumber\\{}
\tau_{ab}=[\tau_{\rm mec}]_{ab}&=&p [\Pi(u)_{ab}+A R_{ab}],
\nonumber\\
\tau_{0a}=[\tau_{\rm th}]_{0a}&=& A(p-\rho) R_{0a},
\end{eqnarray}
in turn re-expressible in terms of the tidal fields ${\mathcal E}(u)$, 
${\mathcal H}(u)$ and ${\mathcal F}(u)$.

In the following sections we are going to write and possibly solve Eqs. 
\eqref{(1.6)} (analytically, or numerically when analytic treatments are very difficult) for
various geometrically meaningful backgrounds.
Let us further note that Eq. \eqref{(1.6)} can also be written as
\begin{equation}
E_{\mu \nu}-B
R_{(\mu}^{\; \; \; \alpha} \;
T_{|\alpha|\nu)} =\kappa  T_{\mu \nu}.
\label{(1.6bis)}
\end{equation}
This equation can be used to re-express $T_{\mu \nu}$ in the exact form
(recalling that $A \equiv 
\frac{B}{\kappa}$)
\begin{equation}
T_{\mu \nu}=\frac{1}{\kappa}E_{\mu \nu}
-{A \over 2}R_{\mu}{}^{\alpha} T_{\alpha \nu}
-{A \over 2}R_{\nu}{}^{\alpha} T_{\alpha \mu},
\end{equation}
By re-inserting it into the left-hand side of Eq. \eqref{(1.6bis)} and recalling that
\begin{eqnarray}
R_{\mu}{}^{\alpha}&=&E_{\mu}{}^{\alpha}+{1 \over 2}\delta_{\mu}^{\; \alpha}R 
-\Lambda \delta_{\mu}^{\; \alpha},
\nonumber\\
R&=&E +2R 
-4 \Lambda  
\end{eqnarray}
that is $R=-E+4\Lambda$, and hence
\begin{equation}
R_{\mu}{}^{\alpha}=E_{\mu}{}^{\alpha}-{1 \over 2}\delta_{\mu}^{\; \alpha} (E-2 \Lambda), 
\end{equation}
we finally obtain the full Einstein tensor to linear order in $A$:
\begin{widetext}
\begin{eqnarray}
\label{eq:squareE}
E_{\mu\nu}&=&\kappa T_{\mu\nu}
+{A \over 2}\left(E_{\mu}{}^{\alpha}+{1 \over 2}\delta_{\mu}^{\; \alpha}R 
-\Lambda \delta_{\mu}^{\; \alpha}\right) E_{\alpha \nu}
+{A \over 2} \left(E_{\nu}{}^{\alpha}+{1 \over 2}\delta_{\nu}^{\; \alpha}R
-\Lambda \delta_{\nu}^{\; \alpha}\right)E_{\alpha \mu}+{\rm O}(A^{2}) 
\nonumber\\
&=& \kappa T_{\mu\nu} +T^A_{\mu\nu},
\end{eqnarray}
where $E \equiv g^{\mu \nu} E_{\mu \nu}$ 
and we have defined
\begin{equation}
T^A_{\mu\nu} \equiv A \left[E_{\mu}^{\; \alpha} \; E_{\alpha \nu}
-{1 \over 2}E_{\mu \nu}(E-2 \Lambda)\right].
\end{equation}
\end{widetext}
The result is then either a $f(R)$ gravity theory  
or a modification of the Einstein's field equations by the addition of an 
extra energy-momentum tensor, completely geometrically motivated and 
small (see Eq. \eqref{eq:squareE}).
The latter is not the main interest in the present 
study, which as stated above, assumes 
finite values of the dimensionless parameter $b$ occurring in $B$   
or $\epsilon$ if one uses the rescaled version of $B$ termed as $A$.
However, when working with a finite value of $b$ will imply excessive 
difficulties, we will also explore the case of infinitesimal $b$.

The case of a constant curvature spacetime is also relevant
\begin{equation}
R^{\alpha\beta}{}_{\mu\nu}=\frac{R}{12}\delta^{\alpha\beta}_{\mu\nu},
\qquad R_{\mu\nu}=\frac{R}{4}g_{\mu\nu},
\end{equation} 
with $R$ constant. Equations \eqref{(1.6)} become then
\begin{equation}
\frac{\Lambda -\frac{1}{4}R}{1+\frac{A}{2}R}g_{\mu\nu}=\kappa T_{\mu\nu}.
\end{equation}
In the case of a perfect fluid this equation is compatible with 
\begin{equation}
p=-\rho=\frac{\Lambda -\frac{R}{4}}{1+\frac{A}{2}R}.
\end{equation}

\section{Modifying perfect fluid spacetimes}

Let us consider the Dunn and Tupper spacetime (see \cite{DT1}, 
and Chapter 12 of \cite{Stephani:2003tm}). This was discovered by
looking for solutions of the Einstein-Maxwell equations for source-free
electromagnetic fields \cite{Tariq}. 
The spacetime metric of this solution reads as
\begin{equation}
ds^{2}=-2 \; du \; dr + u^{-2n}r^{-2m}dy^{2}
+u^{-2m}r^{-2n}dz^{2},
\label{(3.1)}
\end{equation}
where
\begin{equation}
m={(\sqrt{3}-1)\over 4}, \;
n=-{(\sqrt{3}+1)\over 4}.
\label{(3.2)}
\end{equation}
This solution has the property that the principal null congruences
of the electromagnetic field are geodesic, and the corresponding
null tetrad is parallelly propagated along these congruences. It is the 
unique twist-free solution with this property. If one performs the 
coordinate transformation
\begin{eqnarray}
t &=& \sqrt{2ur}, \; x={(m-n)\over 2}\log \left({r \over u}\right),
\nonumber \\
&& y \rightarrow 2^{(m+n) \over 2} \; y, \;
z \rightarrow 2^{(m+n) \over 2} \; z,
\label{(3.3)}
\end{eqnarray}
the metric \eqref{(3.1)} takes the form
\begin{equation}
ds^2=-dt^2+\frac{t^2}{(m-n)^2}dx^2+t^{-2(m+n)}[e^{-2x}dy^2+e^{2x}dz^2], 
\label{(3.4)}
\end{equation}
(for $m\not =n$) 
from which it is clear that the new coordinate system is comoving.
Last, but not least, one investigates the possibility of adapting the
metric \eqref{(3.4)}, so that it represents a perfect-fluid matter
distribution. For this purpose, one can no longer impose the particular
values \eqref{(3.2)}, but a restriction on the admissible values
of $m$ and $n$ (see below) is still necessary in order to obtain a
solution of the Einstein equations.  

Unlike our Secs. I and II, where we needed physical dimensions, 
here the coordinates $t,x,y,z$ are all dimensionless and we may assume that the 
physical coordinates scale with the same constant ${\mathcal L}$,
\begin{equation}
x^{\alpha}_{\rm phys}={\mathcal L} x^\alpha.
\label{(3.5)}
\end{equation}
The metric \eqref{(3.4)} is an exact solution of the Einstein equations in absence of 
cosmological constant and sourced by a perfect fluid 
with four-velocity $u=\partial_t$ 
(i.e., at rest with respect to the space coordinates) and
\begin{equation}
{\mathcal L}^2 \kappa \rho_0= \frac{m^2+mn+n^2}{t^2} ,\qquad 
{\mathcal L}^2 \kappa p_0 = -\frac{4mn}{t^2},
\label{(3.6)}
\end{equation}
provided the dimensionless constants 
$m$ and $n$ satisfy the additional constraint
\begin{equation}
m(2m+1)+n(2n+1)=0.
\label{(3.7)}
\end{equation}

We will conveniently work in the rest of this section with the dimensionless 
coordinates, but restoring the physical ones with the length 
scale ${\mathcal L}$ when necessary.

Note that the conditions $\rho_0>0$ and $p_0\ge 0$ require $mn\le 0$.  
The solution for a dust fluid (i.e., with $p_0= 0$) corresponds to either
$m = 0$ or $n = 0$, but not both of them vanishing since in that 
case the spacetime would be flat. Also the strong energy conditions
\begin{equation}
\rho_0+p_0\ge 0 ,\qquad \rho_0+3p_0\ge 0,
\end{equation}
are always satisfied~\footnote{One should require, however, $m\not= n$, 
as already assumed in Eq. \eqref{(3.4)}. Relaxing this condition is 
possible, but one should revert to the original form of the metric.}.

Equation \eqref{(3.7)} sets the relative dependence of $m$ 
and $n$. Other parametrizations can be found in order to 
satisfy automatically the constraint \eqref{(3.7)} which represents a 
circle in the space of the parameters $m$ and $n$, e.g.,
\begin{equation}
\label{param_alpha}
m=-\frac14 +\frac{1}{2\sqrt{2}}\cos \alpha\,,\qquad n
=-\frac14 +\frac{1}{2\sqrt{2}}\sin \alpha ,
\end{equation}
with $\alpha \in [0,2\pi]$, $\alpha\not =\pi/4$.
The associated \lq\lq sound speed,"
\begin{eqnarray}
v_s(m,n)&=&\sqrt{\frac{p_0}{\rho_0}}=\sqrt{\frac{-4mn}{m^2+mn+n^2}}\nonumber\\
&=& 2\sqrt{\frac{-\frac{n}{m}}{1+\frac{n}{m}+\frac{n^2}{m^2}}},
\end{eqnarray}
is then a constant dependent on the parameters $m$ and $n$ 
(actually it is a function of the ratio $n/m$).
In terms of the parameter $\alpha$ the above relation becomes
\begin{equation}
v_s(\alpha)=2\sqrt{\frac{\sqrt{2}(\sin\alpha+\cos\alpha)-1
-\sin(2\alpha)}{5+\sin(2\alpha)-3\sqrt{2}(\sin\alpha+\cos\alpha)}}.
\end{equation}
It is easy to see that $v_s(n,m)$ vanishes at $n=0$ (or $m\to \infty)$, 
or equivalently at $\alpha=3/4\pi, 7/4\pi$. This velocity reaches 
its maximum value $v_s(m,n)=2$ at $m=-n$, i.e. $\alpha=\pi/4$,  so 
that forcing it to stay in the physical region would further 
restrict the range of allowed parameters. For example
$v_s(m,n)=1$ corresponds to $m/n=(5\pm \sqrt{21})/2$.

Let us shortly review some geometrical properties of the metric 
\eqref{(3.4)} which has not received much attention in the recent literature.  

A Lorentz frame (adapted to $u=\partial_t$) reads
\begin{eqnarray}
\label{frame_u}
&& e_{\hat 0}=u,\quad 
e_{\hat 1}=\frac{(m-n)}{t}\partial_x,
\nonumber\\
&& e_{\hat 2}=e^x t^{m+n}\partial_y,\quad
e_{\hat 3}=e^{-x} t^{m+n}\partial_z. 
\end{eqnarray}

When expressed with respect to the frame \eqref{frame_u} the Riemann tensor components simplify as
\begin{eqnarray}
R_{\hat 0 \hat 2\hat 0 \hat 2} &=& R_{\hat 0 \hat 3\hat 0 \hat 3} 
= -\frac{(m+n)(m+n+1)}{t^2},
\nonumber\\ 
R_{\hat 0 \hat 2\hat 1 \hat 2} &=&-R_{\hat 0 \hat 3\hat 1 \hat 3}= -\frac{(m-n)(m+n+1)}{t^2},
\nonumber\\  
R_{\hat 2 \hat 3\hat 2 \hat 3} &=& -\frac{(m+n)}{t^2},
\nonumber\\ 
R_{\hat 1 \hat 2\hat 1 \hat 2} &=&R_{\hat 1 \hat 3\hat 1 \hat 3}= -\frac{( m+ n-4 m n)}{2t^2} , 
\end{eqnarray}
and are assembled in ${\mathcal E}(u)$, ${\mathcal H}(u)$ and ${\mathcal F}(u)$ as
\begin{eqnarray}
{\mathcal E}(u)_{\hat a\hat b}&=&-\frac{(m+n)(m+n+1)}{t^2} \begin{pmatrix}
0 & 0 & 0 \cr
0 & 1 & 0 \cr
0 & 0 & 1 \cr
\end{pmatrix},
\nonumber\\
{\mathcal H}(u)_{\hat a\hat b}&=&\frac{(m-n)(m+n+1)}{t^2} \begin{pmatrix}
0 & 0 & 0 \cr
0 & 0 & -1 \cr
0 & 1 & 0 \cr
\end{pmatrix},
\nonumber\\
{\mathcal F}(u)_{\hat a\hat b}&=& -\frac{1}{t^2} \begin{pmatrix}
 (m+n)  & 0 & 0 \cr
0 &  \frac{( m+ n-4m n)}{2 } & 0 \cr
0 & 0 &  \frac{( m+ n-4 m n)}{2} \cr
\end{pmatrix}.
\nonumber\\
\end{eqnarray}
Similarly, the nonvanishing frame components of the Ricci tensor are
\begin{eqnarray}
R_{\hat 0 \hat 0}&=&-\frac{2(m+n)(m+n+1)}{t^2} 
\nonumber\\
R_{\hat 1 \hat 1}&=&-\frac{2(m^2-2 m n+n^2+m+n)}{t^2}  
\nonumber\\
R_{\hat 2 \hat 2}&=&R_{\hat 3 \hat 3}=\frac{2(m+n)^2}{t^2}  ,
\end{eqnarray}
and the Ricci scalar is then given by
\begin{equation}
R=\frac{2 (-m-n+8 m n)}{t^2}.
\end{equation}

Moreover, the metric \eqref{(3.4)} is in general of Petrov type I. In fact, 
in a standard Newman-Penrose frame built by using
the Lorentz frame \eqref{frame_u}, with 
\begin{equation}
l=\frac{1}{\sqrt{2}}(e_0+e_1),\quad
n=\frac{1}{\sqrt{2}}(e_0-e_1),\quad
m=\frac{1}{\sqrt{2}}(e_2+i e_3),
\end{equation}
the nonvanishing Weyl scalars are
\begin{eqnarray}
\psi_0&=&-\psi_4=-\frac{(m^2-n^2+m-n)}{t^2},
\nonumber\\
\psi_2&=& -\frac{(m-n)^2}{t^2}.
\end{eqnarray}
The speciality of the metric would imply the relation
\begin{equation}
I^3=27 J^2
\end{equation}
or, introducing the speciality index
\begin{equation}
{\mathcal S}=\frac{I^3}{27J^2}=1,
\end{equation}
where, in the present case with $\psi_1=0=\psi_3$
\begin{equation}
I=3\psi_2^2+\psi_0\psi_4\,,\qquad
J=\psi_2(\psi_0\psi_4-\psi_2^2).
\end{equation}
It is convenient to introduce the (dimensionless) ratio
\begin{equation}
\gamma= -\frac{\psi_2^2}{\psi_0\psi_4}=\left(\frac{ m-n }{3(m+n+1)}\right)^2,
\end{equation}
such that
\begin{equation}
\label{S_in_terms_of_gamma}
{\mathcal S}=\frac{(3\gamma-1)^3}{27\gamma (1+\gamma)^2} .
\end{equation}
Equation \eqref{S_in_terms_of_gamma} shows that the condition of 
being algebraically special (${\mathcal S}=1$)
is approached as soon as $\gamma \to \infty$, or $m=-n-1$.
In that case the spacetime becomes of Petrov type D, where only 
the Weyl scalar survives and equals
\begin{equation}
\psi_2= -\frac{(2n+1)^2}{t^2}.
\end{equation}
Moreover, ${\mathcal S}=0$ at $\gamma=1/3$, that is for
\begin{equation}
m=-\frac{(6+7n)}{(7+4n)}.
\end{equation}

Finally, the geodesic equations (for any causality condition) read 
\begin{eqnarray}
\frac{d^2t}{d\lambda^2} &=& -\frac{t }{(m-n)^2}\left(\frac{dx}{d\lambda}\right)^2
\nonumber\\
&&+(m+n) (P_y^2e^{2x} +P_z^2 e^{-2x}) t^{2 m+2 n-1}
\nonumber\\
\frac{d^2x}{d\lambda^2}&=& -\frac{2}{t}\frac{dt}{d\lambda}\frac{dx}{d\lambda} 
\nonumber\\
&&+(m-n)^2 (P_z^2 e^{-2x}-P_y^2e^{2x})t^{2 m+2 n-2}
\nonumber\\ 
\frac{dy}{d\lambda}&=& P_y e^{2x} t^{2(m+n)} 
\nonumber\\
\frac{dz}{d\lambda}&=& P_z e^{-2x} t^{2(m+n)},
\end{eqnarray}
where $\lambda$ is an affine parameter\footnote{To the best of our 
knowledge this study is absent in the literature.}.
It is immediate to recognize that a particle at rest with respect to the 
coordinates, i.e., with $x=x_0$, $y=y_0$ and $z=z_0$ ($x_0$, $y_0$ and 
$z_0$ constant, implying $P_y=P_z=0$) follows a timelike geodesic with 
$t(\lambda)=c_1\lambda +c_0$. It is worth discussing in detail the 
special case $P_y=P_z=0$ (or $y=y_0$ and $z=z_0$) with the above equations reducing to
\begin{eqnarray}
\frac{d^2t}{d\lambda^2} &=& -\frac{t }{(m-n)^2}
\left(\frac{dx}{d\lambda}\right)^2 ,\nonumber\\
\frac{d^2x}{d\lambda^2} &=& -\frac{2}{t}\frac{dt}{d\lambda}\frac{dx}{d\lambda}.
\end{eqnarray}
The $x$-equation implies then
\begin{equation}
\frac{dx}{d\lambda}=\frac{C_1^2}{t^2},
\end{equation}
[$C_1$ can be assumed as positive without any loss of generality] 
which once inserted in the first one gives
\begin{equation}
\frac{d^2t}{d\lambda^2} = -\frac{C_1^4}{(m-n)^2 t^3}, 
\end{equation}
and can be easily reduced to a first-order equation multiplying 
both sides by $2\frac{dt}{d\lambda}$, 
\begin{equation}
\left(\frac{dt}{d\lambda}\right)^2=\frac{C_1^4}{(m-n)^2 t^2}+ {\rm const}.
\end{equation}
Let us re-name the constant term in the above equation as $C_1^2C_2/4$,
\begin{equation}
\left(\frac{dt}{d\lambda}\right)^2=\frac{C_1^4}{(m-n)^2 t^2}+ \frac{C_1^2C_2}{4} ,
\end{equation}
and introduce the new variable $T=C_1t$, so that
\begin{equation}
\label{dT_dlambda_sq}
\left(\frac{dT}{d\lambda}\right)^2=\frac{1}{(m-n)^2 T^2}+ \frac{C_2}{4} .
\end{equation}
We will assume hereafter $C_2>0$: otherwise Eq. \eqref{dT_dlambda_sq} would 
be valid only in a bounded interval of the temporal coordinate, 
a case which we are not interested in here. 
The last equation can be rewritten as
\begin{equation}
4T^2 \left(\frac{dT}{d\lambda}\right)^2=\left(\frac{dT^2}{d\lambda}\right)^2
=\frac{4}{(m-n)^2}+  C_2T^2 ,
\end{equation}
that is
\begin{equation}
\frac{1}{C_2^2}\left(C_2\frac{dT^2}{d\lambda}\right)^2=\frac{4}{(m-n)^2}
+  C_2T^2 .
\end{equation}
On introducing ${\mathcal T} \equiv \frac{4}{(m-n)^2}+  C_2T^2$ one finds
\begin{equation}
\frac{1}{C_2^2}\left( \frac{d{\mathcal T}}{d\lambda}\right)^2={\mathcal T},
\end{equation}
and hence
\begin{equation}
{\mathcal T}(\lambda)=\left(\pm \frac{C_2}{2}\lambda +C_3  \right)^2 ,
\end{equation}
or
\begin{eqnarray}
&& C_2C_1^2 t^2=\left(\pm \frac{C_2}{2}\lambda +C_3  \right)^2-\frac{4}{(m-n)^2}
\nonumber\\
&=&\left(\pm \frac{C_2}{2}\lambda +C_3  -\frac{2}{m-n}\right)
\left(\pm \frac{C_2}{2}\lambda +C_3  +\frac{2}{m-n}\right).
\nonumber\\ 
\end{eqnarray}
The $\pm$ sign choice in front of $\lambda$ should be set as a plus sign if 
one wants the orbit to be future-oriented, $dt/d\lambda>0$:
\begin{equation}
C_2C_1^2 t^2 
=\left( \frac{C_2}{2}\lambda +C_3  -\frac{2}{m-n}\right)
\left( \frac{C_2}{2}\lambda +C_3  +\frac{2}{m-n}\right). 
\end{equation}
Moreover, if we require  
\begin{equation}
t(0)=0,
\end{equation}
we may then choose the constant in such a way that
\begin{equation}
C_3^2= \frac{4}{(m-n)^2}. 
\end{equation}
Let us define
\begin{equation}
A_\pm (\lambda)=\frac{1}{C_1\sqrt{C_2}}\left( \frac{C_2}{2}\lambda 
+C_3  \pm \frac{2}{m-n}\right).
\end{equation}
We have eventually
\begin{equation}
t(\lambda)=\sqrt{A_+ A_- },
\end{equation}
and hence
\begin{equation}
x(\lambda)=C_4 +\frac{128}{C_2^2(m-n)^3(A_+-A_-)^4}\ln \left(\frac{A_-}{A_+}\right).
\end{equation}
For large values of $\lambda$ we see that 
$A_\pm(\lambda)\to \frac{\sqrt{C_2}}{2 C_1}\lambda$ and 
$x\to C_4$ ($A_+-A_-$ does not depend on $\lambda$), while 
$t\sim \frac{\sqrt{C_2}}{2C_1} \lambda $.
$C_4$ can be therefore identified with 
$x_\infty=\lim_{\lambda \to \infty}x(\lambda)$.

We will specialize our considerations below to 
the case of a dust fluid ($p_0=0$), with
\begin{equation}
m=0,\qquad n=-\frac12 .
\end{equation}
Working at linear order in $A$ and restoring the physical length scale 
${\mathcal L}$ so that in this case
\begin{equation}
\epsilon=\frac{A}{{\mathcal L}^2},
\end{equation}
one modifies this solution to satisfy the new 
equations by changing simply the $tt$ component of the metric as
\begin{equation}
g_{tt}=-1+\epsilon  \frac{4C_p}{5t^2}
\end{equation}
and the energy and pressure of the fluid 
\begin{equation}
\rho=\rho_0 +\epsilon \rho_1 ,\qquad p=p_0+\epsilon  p_1
\end{equation}
with
\begin{equation}
\kappa {\mathcal L}^2 \rho_0=\frac{1}{t^2} ,\qquad p_0=0,
\end{equation}
and
\begin{equation}
\kappa {\mathcal L}^2 \rho_1=\frac{C_\rho}{ t^4},\qquad 
\kappa {\mathcal L}^2 p_1=\frac{C_p}{ t^4},
\end{equation}
with the constraint
\begin{equation}
C_\rho = -\frac12 +C_p .
\end{equation}
Note that the additional piece (linearizing in $\epsilon$, i.e. in $A$) 
results in the following tensor:
\begin{eqnarray}
&&\kappa {\mathcal L}^4 R_{(\mu}^{\; \; \alpha} \; T_{|\alpha|\nu)}=-{1 \over 2  t^{2}}
\delta_{\mu}^{0} \; \delta_{\nu}^{0}
\nonumber\\
&&+{\epsilon \over  t^{4}}{\rm diag} 
\left[-{(-5+42C_{p})\over 20 t^{2}},2C_{p},C_{p}{e^{-2x}\over 2t},
C_{p}{e^{2x}\over 2t} \right].
\nonumber\\
\end{eqnarray}
If instead of coupling the Ricci tensor to the energy-momentum tensor one uses the
Einstein tensor (see Appendix) the result is simply
\begin{equation}
\kappa {\mathcal L}^4  E_{(\mu}^{\; \; \alpha} \; T_{|\alpha|\nu)}=-{1 \over   t^{4}}
\left[1+\epsilon {(-5+12 C_{p})\over 10 t^{2}}\right]\delta_{\mu}^{0} \;
\delta_{\nu}^{0}.
\end{equation}

Interestingly, one can look for exact solutions also in the general case, i.e., 
not considering the linear expansion in $\epsilon$.
In the case $m=0$, $n=-1/2$ discussed above, still modifying only the 
$tt$ metric component as
\begin{equation}
g_{tt}=-1+\epsilon f_{tt},
\end{equation}
together with energy density and pressure
\begin{equation}
\kappa {\mathcal L}^2 \rho(t) = \frac{1}{t^2}+\epsilon \kappa 
{\mathcal L}^2\rho_1(t)\,,\qquad p(t) = \epsilon p_1(t),  
\end{equation} 
and the exact solution is 
\begin{eqnarray}
\label{ex_sol}
f_{tt}&=& \frac{1}{\epsilon}+\frac{t^{3/4}}{C_1- \frac{(11 
\epsilon +12 t^2) t^{3/4}}{33}}
\nonumber\\
\kappa {\mathcal L}^2 \rho_1(t) &=& -\frac{(\epsilon-t^2)}{ t^2 \epsilon^2} 
\nonumber\\
\kappa {\mathcal L}^2 p_1(t) &=& -\frac{1}{\epsilon^2},
\label{(3.54)}
\end{eqnarray}
where $C_1$ is an integration constant and $\epsilon$ is not necessarily small  
(both $t$ and $\epsilon$ are dimensionless).
In this exact solution, Eq. \eqref{ex_sol}, negative pressure and sign-changing 
energy density during the evolution are evident, confirming the expectations 
of the linearized model for the presence of dark or exotic matter in a universe 
modeled as in this toy model.
 
However, finding -as is this case- an explicit, exact solution is never a trivial task.
It requires always some care, even when using
algebraic manipulation systems like MapleTM and MathematicaTM.
As a first attempt one may try to use  the same symmetries of the
background metric and then proceed by relaxing some hypothesis.
In the present case we have been looking for simple conditions on the fluid
source of the spacetime, like constant energy density and/or pressure with
a minimal backreaction on the modified metric. However, this may not be
enough in general, and one should then analyze the system of coupled
equations, looking for some simplifications.
The risk is to end up with a purely mathematical solution, devoid of physical
meaning, or that any improvement is reached only by trial and error,
without a clear understanding of the intermediate steps. 

\section{Modifying spherically symmetric static spacetimes sourced by a perfect fluid}

In order to investigate the role of the curvature-energy coupling discussed above the 
simplest arena is that associated with an internal solution of the Schwarzschild spacetime.
The latter is a spherically symmetric spacetime, with metric written in the form
\begin{equation}
ds^2=g_{tt}dt^2+g_{rr} dr^2 +r^2(d\theta^2+\sin^2\theta d\phi^2),
\label{(4.1)}
\end{equation}
(with $g_{tt}$ and $g_{rr}$ depending only on $r$) 
sourced by a perfect fluid with a constant energy density and in absence of cosmological constant.
Let us introduce the notation
\begin{equation}
{\mathcal F}(x,y) \equiv \sqrt{1-\frac{x^2}{y^2}},
\label{(4.2)}
\end{equation}
with ${\mathcal F}$ dimensionless. 

The interior Schwarzschild solution, for example, corresponds to
\begin{eqnarray}
g_{tt}^{\rm is}&=&-\frac14 [3{\mathcal F}(r_s,R)-{\mathcal F}(r,R)]^2 
\nonumber\\
g_{rr}^{\rm is}&=&{\mathcal F}^2(r,R)  
\nonumber\\
\rho_0&=&\frac{3}{\kappa R^2} 
\nonumber\\
p_0&=& \rho^{\rm is} \frac{{\mathcal F}(r,R)
-{\mathcal F}(r_s,R)}{3{\mathcal F}(r_s,R)-{\mathcal F}(r,R)} ,
\label{(4.3)}
\end{eqnarray}
where $r_s=2M$ denotes the Schwarzschild radius and $R=\sqrt{r_{\rm body}^3/r_s}$ is a 
length scale built with the radius of the interior  \lq\lq body." 
In order to shorten equations we will introduce the notation
\begin{equation}
{\mathcal F}_r={\mathcal F}(r,R),\qquad 
{\mathcal F}_s={\mathcal F}(r_s,R).
\end{equation}

The $\epsilon$-modifications 
to this rather simple solution are not simple at all, and in general one is left only with 
the numerical integration of the associated equations.
One can look at linear perturbations of the interior Schwarzschild solution, i.e.
\begin{eqnarray}
g_{tt}&=& g_{tt}^{\rm is}+\epsilon f_{tt} 
\nonumber\\
g_{rr}&=& g_{rr}^{\rm is}+\epsilon f_{rr}  
\nonumber\\
\rho&=&\rho_0+\epsilon \rho_1 
\nonumber\\
p&=&p_0+\epsilon p_1 ,
\label{(4.4)}
\end{eqnarray}
assuming spherical symmetry (all $B$-corrections depend only on the radial 
variable and $B$ itself is given by $B=\kappa R^2 \epsilon$). 
Formally, one can introduce a vector notation for the unknown functions 
\begin{equation}
{\mathbf X}=[X_1,X_2,X_3,X_4],  
\end{equation}
with $X_1=f_{tt}$, $X_2=f_{rr}$, $X_3=\rho_1$  and $X_4=p_1$
and write down a system of coupled linear equations
\begin{equation}
\frac{d}{dr}X_i=A_{ij} X_j +C_i ,
\end{equation}
with $A_{ij}$ and $C_i$ depending on $r$ and
whose explicit expression does not involve an equation for $d\rho_1/dr$ 
(the perturbation equations are actually three). It is summarized by 
the only nonvanishing components listed below
\begin{eqnarray}
A_{11} &=& -\frac{2 r}{ R^2 {\mathcal F}_r ({\mathcal F}_r-3 {\mathcal F}_s)},
\nonumber\\  
A_{12}&=& -\frac{(3 {\mathcal F}_s-{\mathcal F}_r) {\mathcal F}_r (3{\mathcal F}_s 
{\mathcal F}_r-3 {\mathcal F}_r^2+2)}{4r},  
\nonumber\\
A_{22} &=& -\frac{({\mathcal F}_r R-2 r) ({\mathcal F}_r R+2 r)}{ R^2 F_r^2 r},
\nonumber\\ 
A_{42} &=& \frac{3 {\mathcal F}_s {\mathcal F}_r ({\mathcal F}_r^2-1) 
(3 {\mathcal F}_s {\mathcal F}_r-3 {\mathcal F}_r^2+2)}{\kappa r^3 
(3 {\mathcal F}_s-{\mathcal F}_r)^2},
\nonumber\\ 
A_{23}&=& \frac{r\kappa}{{\mathcal F}_r^4},
\nonumber\\ 
A_{43}&=& \frac{({\mathcal F}_r^2-1)}{ {\mathcal F}_r r 
(3 {\mathcal F}_s-{\mathcal F}_r)},
\nonumber\\ 
A_{14}&=& -\frac{(3 {\mathcal F}_s-{\mathcal F}_r)^2 r\kappa}{4 {\mathcal F}_r^2}, 
\nonumber\\ 
A_{44}&=& -\frac{(3 {\mathcal F}_s+{\mathcal F}_r) r}{R^2(3 
{\mathcal F}_s-{\mathcal F}_r) {\mathcal F}_r^2} , 
\end{eqnarray}
and
\begin{eqnarray}
C_1 &=& \frac{9(2{\mathcal F}_s-{\mathcal F}_r) 
({\mathcal F}_s-{\mathcal F}_r) r}{4 R^2 {\mathcal F}_r^2}, 
\nonumber\\  
C_2 &=& -\frac{9 r}{ R^2 {\mathcal F}_r^3 (3 {\mathcal F}_s-{\mathcal F}_r)},
\nonumber\\  
C_4 &=& \frac{18 {\mathcal F}_s^2 ({\mathcal F}_r^2-1)^2 }{\kappa 
(3 {\mathcal F}_s-{\mathcal F}_r)^2 {\mathcal F}_r^2 r^3}.
\end{eqnarray}

\begin{figure}
\includegraphics[scale=0.35]{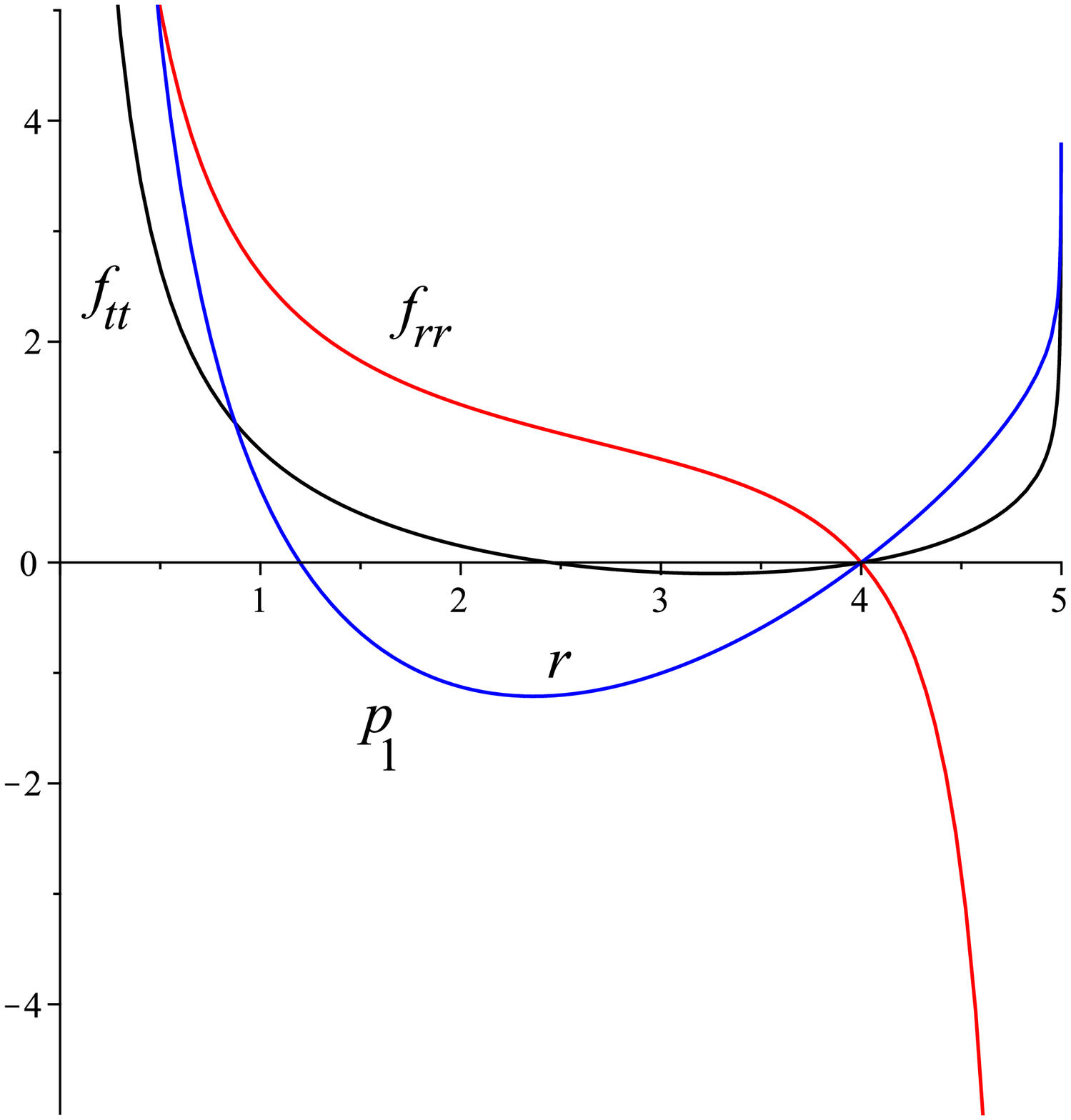}
\includegraphics[scale=0.35]{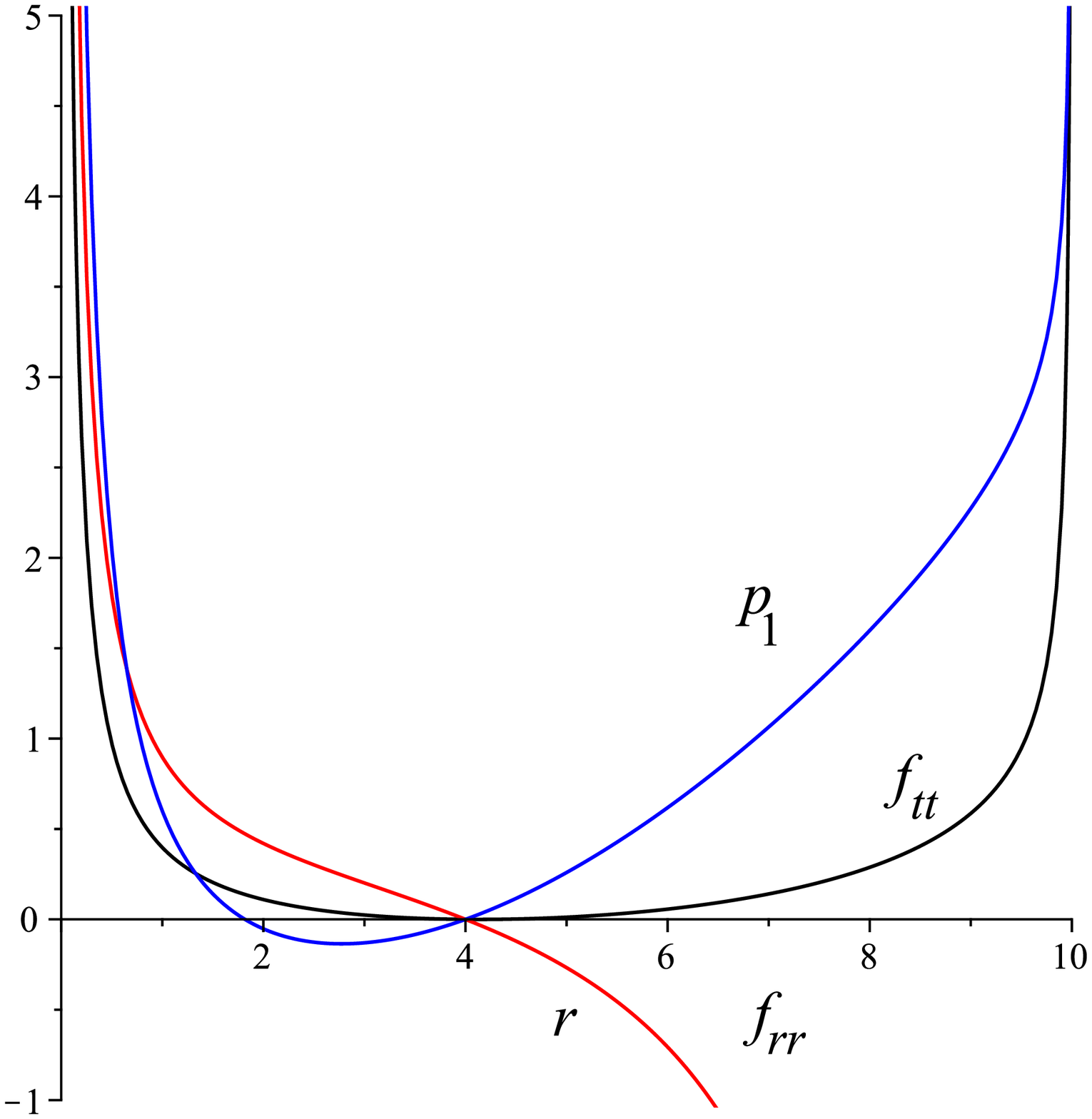}
\caption{\label{fi1_schw} The metric perturbations $f_{tt}$, $f_{rr}$ 
and $p_1(r)$ (in units of $\kappa R^2$)
are shown as functions of $r$ in the special case of
$\rho_B=1$ (in units of $\kappa R^2$) and $r_s=R/2$ and $R=5$ 
(upper plot) and $R=10$ (lower plot). 
Initial conditions are chosen in 
$r=4$ as $f_{rr}(4)=f_{tt}(4)=p_1(4)=0$. Because of the coordinate 
dependence of the various quantities this plot is illustrative but 
qualitative for what concerns its physical content.
}
\end{figure}

Of course, in this case the 
integration of the full system can only be carried out numerically and an example is given
in Fig. \ref{fi1_schw}.

The form of the system is such that the evolution equation for 
$\rho_1$ is implicit in the compatibility of the system. Consequently 
$\rho_1$=constant is a natural choice to cast the perturbation 
equations in their normal form.
This is what has been done in the case of Fig. \ref{fi1_schw}. 
[Actually, it is worth mentioning that an exact solution for $f_{rr}$ 
can also be found in this case. We will not display it here because 
of its length and because it does not add much to the present discussion.] 
Numerical integration shows that the perturbed pressure can be negative during its 
evolution, leaving also in this case the possibility open for the birth 
of dark or exotic matter during the evolution.

\section{Modifying FLRW spacetimes sourced by a perfect fluid and with 
nonvanishing cosmological constant}

Let us consider a FLRW background spacetime with metric written in 
spherical-like coordinates, sourced by a perfect fluid and in presence 
of a nonvanishing cosmological term:
\begin{equation}
ds^2=-dt^2+a^2(t)\left[\frac{dr^2}{(1-kr^2)}+f^{2}(r)(d\theta^2+\sin^2\theta d\phi^2)\right],
\end{equation}
with $k=[-1,0,1]$, corresponding to a closed ($k=-1$), spatially flat ($k=0$), 
open ($k=1$) $t$=constant $3$-spaces and 
\begin{equation}
f(r)=\left\{ \begin{array}{ll}
\sinh r & k=-1\cr
r & k=0 \cr
\sin r & k=1. \cr
\end{array}
\right.
\end{equation}
Here $t$ and $a$ have the dimensions of a length while $r$, $\theta$ and $\phi$ are dimensionless.

The perfect fluid, source of this spacetime, is assumed at rest with respect 
to the chosen coordinate system, i.e., it is associated with a four-velocity 
field $u=\partial_t$ and its energy-momentum tensor reads in general as
\begin{equation}
T_{\mu\nu}=\rho_0(t) u_\mu u_\nu +p_0(t) \Pi(u)_{\mu \nu}\,,  
\end{equation}
with
\begin{equation}
\Pi(u)_{\mu\nu}=g_{\mu\nu}+u_\mu u_\nu ,
\end{equation}
and the Einstein's field equations are given by
\begin{eqnarray}
\left(\frac{\dot a}{a}\right)^2&=&\frac{\Lambda}{3} 
-\frac{k}{a^2}+\frac{\kappa}{3}\rho_0 ,
\nonumber\\
\frac{\ddot a}{a}&=&\frac{\Lambda}{3}-\frac{\kappa }{2}
\left(\frac13 \rho_0+p_0\right),
\end{eqnarray}
plus the compatibility condition
\begin{equation}
\dot \rho_0=-3 \frac{\dot a}{a}(\rho_0+p_0).
\end{equation}

The modified equations become
\begin{eqnarray}
\frac{\ddot a}{a}&=& \frac13 \frac{2A\kappa p (\kappa  \rho+ \Lambda)
-\kappa(\rho+3p)+2\Lambda}{2-A\kappa (\rho+p)+2A^2\kappa^2 p \rho} ,
\nonumber\\
\frac{\dot a^2}{a^2} &=&  \frac{  [2\kappa^2\rho p -\kappa\Lambda(3 \rho + p)] 
A+2\kappa\rho+2\Lambda }{3 [2-A\kappa ( p +\rho) 
+2 A^2\kappa^2\rho p ]}  -\frac{k}{ a^2} . 
\nonumber\\ 
\end{eqnarray}

Let us consider for simplicity the spatially flat case $k=0$ and let us assume $\Lambda=0$.
The solution of the Einstein's field equation is termed 
Friedmann-Lemaitre solution and is given by
\begin{eqnarray}
\label{FL_sol}
\kappa t_0^2 \rho_0&=&\frac{4 t_0^2}{3\gamma^2 t^2},
\nonumber\\
p_0&=&(\gamma-1)\rho_0,
\nonumber\\
a(t)&=& t_0 \left(\frac{t}{t_0}\right)^{\frac{2}{3\gamma}},
\end{eqnarray}
where $\gamma$ is dimensionless parameter and $t_0$ is an arbitrary 
length scale\footnote{The choice $t_0=1$ makes the form of the solution 
\eqref{FL_sol} more familiar.} associated with $a(t)$.
When $\gamma=1$ it becomes the Einstein-de Sitter Universe solution.
It is convenient to introduce the rescaled, dimensionless time variable,
\begin{equation}
T=\frac{t}{t_0}.
\end{equation}
Looking for perturbative solutions at the first order in 
$B=\kappa L^2 \epsilon$ ($\epsilon$ dimensionless; in this way the perturbed 
quantities have the same dimensions of the corresponding original ones), i.e.,
\begin{eqnarray}
\rho&=&\rho_0+\epsilon \rho_1,
\nonumber\\
p&=&p_0+\epsilon  p_1
\nonumber\\
a&=&a_0+\epsilon  a_1,
\end{eqnarray}
it is straightforward to identify the following solution
\begin{eqnarray}
\kappa t_0^2 \rho_1(t) &=&  - \frac{8(7\gamma-6)}{9 \gamma^3 (3\gamma-2) 
T^4}-\frac{2 p_1 \kappa t_0^2}{(2+\gamma)} + T^{-\frac{(2+\gamma )}{\gamma}},  
\nonumber\\
p_1(t)&=& p_1,
\nonumber\\
\frac{a_1(t)}{t_0}&=&C_1 T^{\frac{2}{3 \gamma}} -\frac{ \gamma \kappa t_0^2 p_1}{ 4 
(2+\gamma)}T^{\frac{2(1+3\gamma)}{3\gamma} }
\nonumber\\
&+&\frac{\gamma^2}{4(\gamma-2)} T^{-\frac{4}{ 3 \gamma }+1} 
-\frac{2(\gamma-1) (\gamma -2) }{9 (3\gamma -2)\gamma^3} T^{\frac{2}{ 3\gamma}-2}, 
\nonumber\\
\end{eqnarray}
where $C_1$ in an integration constant. Note that here 
we have been looking for solutions with $p_1(t)$ constant. This simplifying 
condition (which is enough for the purposes of the present discussion) 
can be eventually relaxed.
Let us assume $\gamma=1$ and $C_1=0$, for a practical purpose. We find
\begin{eqnarray}
\kappa t_0^2 \rho_1(t) &=&-\frac{8}{9 T^4}-\frac23 t_0^2\kappa p_1+\frac{1}{T^3},
\nonumber\\
p_1(t)&=&  p_1,
\nonumber\\
\frac{a_1(t)}{t_0} &=& -\frac{\kappa t_0^2 p_1}{12} T^{\frac83}-\frac14  T^{-\frac13} ,
\end{eqnarray}
showing that $ \rho_{(1)}(t)\to -\frac{2 p_1}{3}$ 
asymptotically, whereas $a_1(t)$ diverges in general.
The special case $p_1=0$ avoids such a divergence and gives an asymptotic 
damping of the perturbation, i.e.,
\begin{eqnarray}
\kappa t_0^2 \rho_{(1)}(t) &=&-\frac{8}{9 T^4}+\frac{1}{T^3},
\nonumber\\
p_1(t)&=& 0,
\nonumber\\
\frac{a_1(t)}{t_0} &=& -\frac14  T^{-\frac13} ,
\label{(5.12)}
\end{eqnarray}
to be compared with the unperturbed values (for $\gamma=1$), 
\begin{equation}
\kappa t_0^2 \rho_0=\frac{4}{3T^2},\qquad p_{(0)}=0,\qquad a_0(t)=t_0 T^{\frac{2}{3}}.
\end{equation}
We have then
\begin{eqnarray}
\label{sol_ein_FL}
\kappa t_0^2 \rho&=&\frac{4}{3T^2}+\epsilon \left(-\frac{8}{9 T^4}+\frac{1}{T^3}\right),
\nonumber\\
\frac{a(t)}{t_0}&=& T^{\frac{2}{3}}-  \frac{\epsilon}4  T^{-\frac13}.
\end{eqnarray}
Therefore, the perturbation  changes both  the background geometry and 
the mass-energy content of the spacetime. Moreover, 
$\rho_1(t)$  may change sign during the evolution and 
it is therefore not an arbitrary speculation to admit that a minimal modification of 
the Einstein's field equation may allow for the theoretical existence of 
dark or exotic matter in some spacetime region.
\begin{figure}
\includegraphics[scale=0.35]{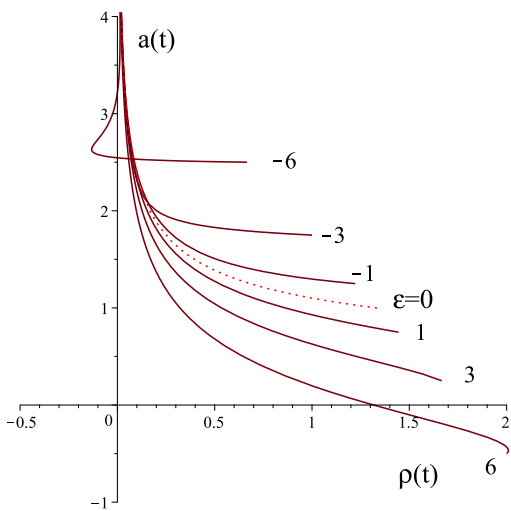}
\caption{\label{fig2} The behavior of $a(t)$ (in units of $t_0$) vs 
$\rho(t)$ (in units of $\kappa t_0^2$) of the solution \eqref{sol_ein_FL} 
is shown for different values of $\epsilon=0,\pm 1, \pm 3, \pm 6$ showing 
the displacement of the perturbed quantities from the unperturbed ones.}
\end{figure}

\section{Concluding remarks}

We have explored the features of a toy model where, in the Einstein equations, 
the right-hand side is modified by the addition of a term
proportional to the symmetrized partial contraction of the Ricci
tensor with the energy-momentum tensor, while the left-hand
side remains equal to the Einstein tensor. 
Indeed, one can modify the Einstein's field equations in a 
number of ways, grounded on geometrical reasons or on physical reasons. 
Our choice of including an \lq\lq R-T" correction is in between but, for 
the purpose of the present study, any particular choice is valid. Thinking of 
\lq\lq small corrections" we have argued that the coupling constant in the 
\lq\lq R-T" term might have a quantum origin, by virtue of the    
existence of a natural length scale given by the Planck length. 
This remark, supplemented by dimensional analysis, 
shows that such a term yields a correction
linear in $\hbar$ to the classical term, that is instead just proportional
to the energy-momentum tensor. 
A nice feature of this model is that it coincides with
general relativity in vacuum and can be related to various $f(R)$ 
theories already studied in the recent literature.  

Motivated by the analysis of the subsequent corrections on the background 
geometry and the background energy-momentum tensor source of the spacetime 
curvature, we have studied linear perturbations by using as unperturbed situation 
some special, non-vacuum exact solutions.
These are the Dunn and Tupper metric, the interior Schwarzschild solution 
and a Friedmann-Lemaitre cosmological solution, besides some general 
considerations concerning the simple case of constant curvature spacetimes.

All the studied situations are interesting and the Friedmann-Lemaitre case is 
also illuminating: it is far from being an arbitrary conjecture that the   
dark or exotic matter may form in some spacetime region, even with a small temporal duration.
This is made clear by our Eqs. \eqref{(2.7)}, \eqref{(2.12)},
\eqref{(3.54)} and \eqref{(5.12)}. Instead of postulating new forms
of matter or new gravitational lagrangians, we have allowed for a novel way of
coupling gravity, i.e., its geometrical description, to matter 
fields, showing that we might need both conceptual
ingredients at once in order to overcome the apparent shortcomings of general 
relativity on large scales. In other words, the net separation of geometry 
(curvature tensors) and physics (matter energy-momentum tensor), which is 
implicit in the Einstein equations, is \lq\lq called into question" in 
the toy model presented here, and modified by the addition of a direct 
(i.e., the simplest possible one) coupling among these two ingredients.  

However, since the right-hand side of our field field equations
(see also \eqref{(A1)}) is tensorial but not variational, the model we 
have introduced and discussed in this paper remains
a toy model, unless one can find a stronger foundation for field equations
whose right-hand side is not variational. This open problem deserves further
attention, since not all partial differential equations of interest are variational
(see, e.g., Ref. \cite{Anco}).
 
\section*{Acknowledgements}
The authors are grateful to Dipartimento di Fisica ``Ettore Pancini'' 
for hospitality and support. DB thanks ICRA and ICRANet for partial support, and MaplesoftTM
for providing a complementary license of MAPLE2020.

\begin{appendix}
\section{Nonlinear coupling of Einstein's tensor to the energy-momentum tensor}

The field equation (1.4) that we have postulated results from considering a nonlinear 
coupling of gravity to matter, and hence suggests considering also the following
alternative:
\begin{equation}
E_{\mu\nu}=R_{\mu \nu}-{1 \over 2}g_{\mu \nu}R+\Lambda g_{\mu\nu} 
=\kappa T_{\mu \nu}+B E_{(\mu}^{\; \; \alpha}
\; T_{|\alpha|\nu)}.
\label{(A1)}
\end{equation}
Now we write explicitly the symmetrization on the right-hand side, finding therefore
\begin{equation}
\left(\delta_{\nu}^{\alpha}-{B \over 2}T_{\nu}^{\alpha}\right)E_{\alpha \mu}
-{B \over 2}T_{\mu}^{\alpha} \; E_{\alpha \nu}=\kappa T_{\mu \nu}.
\label{(A2)}
\end{equation}
This form of the field equation suggests defining the tensor
\begin{equation}
U_{\nu}^{\alpha} \equiv \delta_{\nu}^{\alpha}-{B \over 2}T_{\nu}^{\alpha},
\label{(A3)}
\end{equation}
whose inverse $W_{\beta}^{\nu}$ should fulfill the condition
\begin{equation}
U_{\nu}^{\alpha} \; W_{\beta}^{\nu}=\delta_{\beta}^{\alpha}.
\label{(A4)}
\end{equation}
At this stage, bearing in mind the definition (A3), we multiply both sides
of Eq. (A2) by $W_{\beta}^{\nu}$ and we sum over repeated indices. Hence 
we find, exploiting the symmetry of Einstein's tensor,
\begin{equation}
E_{\mu \beta}-{B \over 2} T_{\mu}^{\alpha} \; E_{\alpha \nu} \;
W_{\beta}^{\nu}=\kappa T_{\mu \nu} \; W_{\beta}^{\nu}.
\label{(A5)}
\end{equation}
We can point out that, upon inserting (A3) into the condition (A4) one finds
the recursive algorithm
\begin{eqnarray}
W_{\beta}^{\alpha}&=& \delta_{\beta}^{\alpha}
+{B \over 2}T_{\nu}^{\alpha} \; W_{\beta}^{\nu}
=\delta_{\beta}^{\alpha}+{B \over 2} T_{\nu}^{\alpha}
\left(\delta_{\beta}^{\nu}+{B \over 2}T_{\gamma}^{\nu}
\; W_{\beta}^{\gamma}\right) 
\nonumber \\
&=& \delta_{\beta}^{\alpha}+{B \over 2} T_{\beta}^{\alpha}
+\left({B \over 2}\right)^{2}T_{\nu}^{\alpha} \; T_{\gamma}^{\nu}
W_{\beta}^{\gamma}=... \; .
\label{(A6)}
\end{eqnarray}
If the dimensionless parameter $b$ introduced in (1.5) approaches $0$, we 
can therefore deal with finitely many powers of the energy-momentum tensor
in Eq. (A5), by truncating the sum of terms on the right-hand side of
Eq. (A6).

\end{appendix}

\end{document}